\documentclass[twocolumn]{aastex631}

\usepackage{multirow}

\shorttitle{Classifying optical (out)bursts in cataclysmic variables}
\shortauthors{I\l{}kiewicz et al.}

\begin{document}

\title{Classifying optical (out)bursts in cataclysmic variables: the distinct observational characteristics of dwarf novae, micronovae, stellar flares and magnetic gating}

\author[0000-0002-4005-5095]{Krystian I\l{}kiewicz}
\affiliation{Astronomical Observatory, University of Warsaw, Al. Ujazdowskie 4, 00-478 Warszawa, Poland}
\affiliation{Centre for Extragalactic Astronomy, Department of Physics, Durham University, DH1 3LE, UK}

\author[0000-0001-5387-7189]{Simone Scaringi}
\affiliation{Centre for Extragalactic Astronomy, Department of Physics, Durham University, DH1 3LE, UK}

\author[0000-0002-0146-3096]{Martina Veresvarska}
\affiliation{Centre for Extragalactic Astronomy, Department of Physics, Durham University, DH1 3LE, UK}

\author[0000-0002-5069-4202]{Domitilla De Martino}
\affiliation{INAF–Osservatorio Astronomico di Capodimonte, salita Moiariello 16, I-80131, Napoli, Italy}

\author[0000-0001-7746-5795]{Colin Littlefield}
\affiliation{Department of Physics, University of Notre Dame, Notre Dame, IN 46556, USA}

\author[0000-0002-1116-2553]{Christian Knigge}
\affiliation{Department of Physics and Astronomy. University of Southampton, Southampton SO17 1BJ, UK}

\author{John A. Paice}
\affiliation{Centre for Extragalactic Astronomy, Department of Physics, Durham University, DH1 3LE, UK}

\author[0009-0007-6825-3230]{Anwesha Sahu}
\affiliation{Department of Physics, University of Warwick, Gibbet Hill Road, Coventry CV4 7AL, UK}

\begin{abstract}

Cataclysmic variables can experience short optical brightenings, which are commonly attributed to phenomena such as dwarf novae outbursts, micronovae, donor flares or magnetic gating bursts. Since these events exhibit similar observational characteristics, their identification has often been ambiguous. In particular, magnetic gating bursts and micronovae have been suggested as alternative interpretations of the same phenomena. Here we show that the timescales and energies separate the optical brightenings into separate clusters consistent with their different classifications.  This suggest that micronovae and magnetic gating bursts are in fact separate phenomena. Based on our finding we develop diagnostic diagrams that can distinguish between these bursts/flares based on their properties.   We demonstrate the effectiveness of this approach on observations of a newly identified intermediate polar, CTCV~J0333-4451, which we classify as a magnetic gating system. CTCV~J0333-4451 is the third high spin-to-orbital period ratio intermediate polar with magnetic gating, suggesting that these bursts are common among these rare systems.

\end{abstract}


~\\

\section{Introduction} \label{sec:intro}

Cataclysmic variables (CVs) are binary stars in which an accreting white dwarf accretes material typically from a main sequence mass donor. CVs can exhibit various forms of short optical bursts. Among the best-studied bursts in CVs are dwarf novae. Dwarf nova eruptions are caused by a thermal-viscous instability in the accretion disc \citep{2001NewAR..45..449L}. Based on the properties of the bursts, dwarf novae are divided into several subtypes \citep[e.g.][]{2003cvs..book.....W}. However, their duration and amplitude depend mainly on the size of the accretion disc. In particular, their duration can range from a few days \citep[e.g.][]{2012ApJ...747..117C} to years \citep[e.g.][]{2023ApJ...953L...7I}.

In CVs with a magnetic white dwarf the accretion disc is truncated. This makes dwarf novae outbursts less likely to occur in magnetic systems compared to non-magnetic CVs \citep{2017A&A...602A.102H}.  In these magnetic systems, two other kinds of bursts  seem more likely: micronovae and magnetic gating bursts. Micronovae occur on timescales of hours, show energies 10$^{-6}$ times smaller compared to classical novae, and have outburst shapes similar to Type~I~X-ray bursts in accreting neutron stars \citep{2022Natur.604..447S}. The proposed mechanism behind micronovae are localized thermonuclear runaways in magnetically confined accretion streams \citep{2022MNRAS.514L..11S}. Another explanation for micronovae could be magnetic reconnection events in the magnetic disc \citep{2022MNRAS.512.1924S}. On the other hand, in the magnetic gating model, accretion is halted by the white dwarf magnetic field until enough pressure builds up in the accretion disc allowing for a short burst of accretion \citep{2010MNRAS.406.1208D,2012MNRAS.420..416D,2017Natur.552..210S}. While magnetic gating is a widely accepted phenomenon, the reality of micronovae is still under question. In particular, magnetic gating bursts have been proposed as an alternative interpretation of the claimed micronovae \citep{2022A&A...664A...7H}. 

Among the most rare short brightenings observed in CVs are stellar flares originating from the mass donor \citep{2021MNRAS.504.4072R}. They are expected to have similar properties to flares and superflares from single stars. However, they are rarely observed in CVs. This is likely because the mass donor is tidally locked, making it spin rapidly, and rapidly rotating main sequence stars are unlikely to show flaring activity \citep{2020MNRAS.497.2320R}.

Here we explore observational properties of short optical bursts in CVs. Based on the studied systems we discover that the short bursts observed in CVs fall into separate clusters based on their burst energies and timescales. In particular, we show that magnetic gating systems and micronovae display distinct sets of characteristics, implying different physical mechanisms are at play in these bursts. Based on our findings, we advocate the use of diagnostic diagrams to identify the nature of a given burst. We employ the newly proposed method to classify bursts in the newly identified intermediate polar CTCV~J0333-4451 (hereafter J0333). We show that J0333 falls in the cluster of magnetic gating systems.

\section{Observations and Literature Data} \label{sec:obs}

We employed observations of J0333 made by \textit{Transiting Exoplanet Survey Satellite} \citep[\textit{TESS}]{2015JATIS...1a4003R}. The observations were carried out during \textit{TESS} sectors 30 and 31 (23 September -- 18 November 2020) with a 120s cadence. The data was processed with the Science Processing Operations Center (SPOC) pipeline \citep{2016SPIE.9913E..3EJ}.

\begin{figure*}
\includegraphics[width=1.0\hsize]{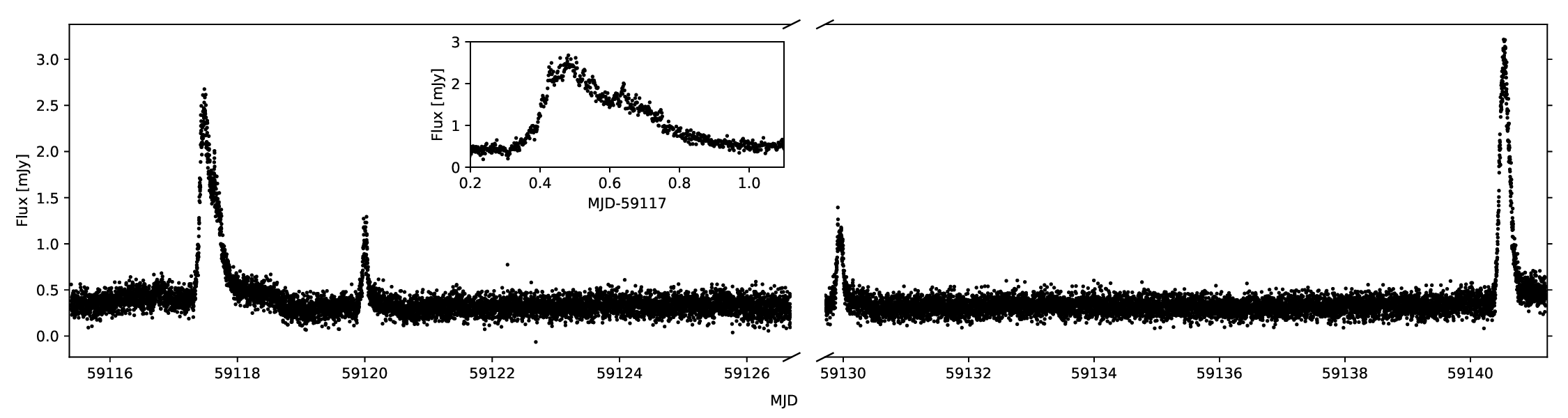}
\includegraphics[width=1.0\hsize]{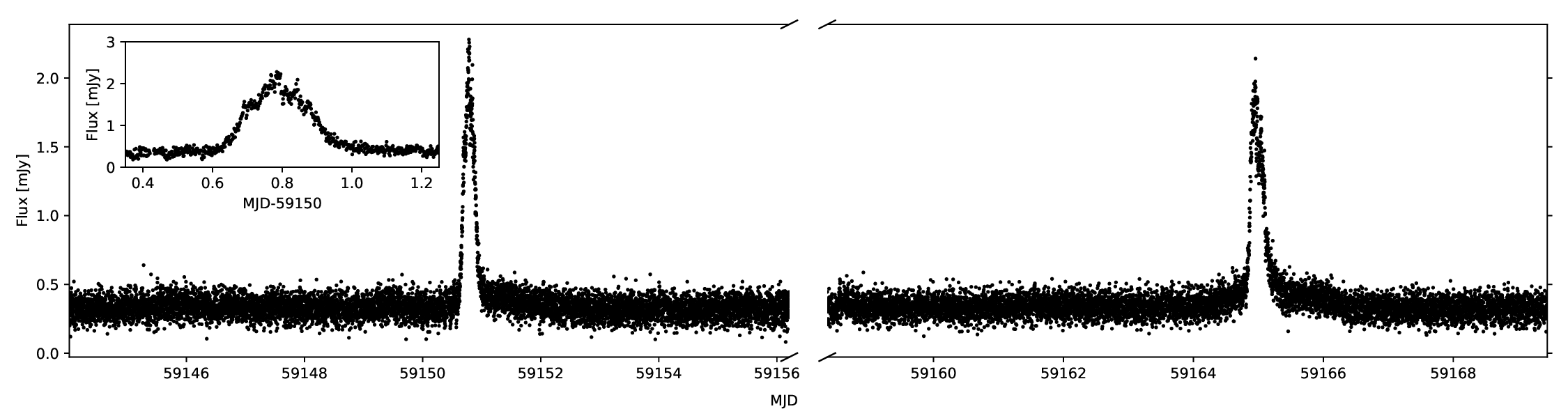}
\caption{Sectors 30 (top) and 31 (bottom) of the $TESS$ observations of J0333 with flux scale calibrated to the ASAS-SN $g$ filter.The top inset figure show a zoom on an asymmetrical burst, while the bottom inset figure shows a symmetrical burst.}
\label{fig:J0333_tess}
\end{figure*}

In order to flux-calibrate \textit{TESS} observations we followed a method employed by \citet{2022Natur.604..447S}. Namely, we found nearly simultaneous observations of \textit{TESS} and ASAS-SN. Then, we fitted a linear relationship between the \textit{TESS} SAP flux in electrons per second and ASAS-SN flux in $g$ band. We then used this relationship to scale the \textit{TESS} data to fluxes in Jy.  The J0333 light curve is presented in Fig.~\ref{fig:J0333_tess}. We employed the same flux calibration method with \textit{TESS} data of J0333 and observations of other objects.

Here we consider properties of short bursts observed in CVs, i.e. bursts with duration of order of days or shorter. Due to the short timescales of the bursts, we limited the comparison to bursts observed with  $Kepler$ and $TESS$, where the cadence is higher and more consistent compared to ground based observations. Our analysis is intended for CVs with short orbital periods ($<$10~hours) and will have limited application to CVs with evolved donors. This is because evolved stars can experience stellar flares that have energies similar to the energies of micronovae \citep[see e.g. fig.~9 of][]{2021ApJS..253...35T}.  However, the only CV with short bursts and a long orbital period observed by either $TESS$ or $Kepler$ is V2487~Oph, with the distance to the object being too uncertain for a meaningful analysis \citep{2022MNRAS.512.1924S}. In order to estimate the energies of bursts in each system we assumed distances based on a \textit{Gaia} DR3 parallax \citep{2021AJ....161..147B}. We also exclude classical novae from our analysis, even though they can occur on short timescales \citep[e.g.][]{2023MNRAS.521.5453S}. This is because classical novae have luminosities several orders of magnitude larger compared to other bursts in CVs.  We assume that interstellar reddening to all of the systems is negligible. This is consistent with extinction A$_g<$0.4~mag estimated with a 3D reddening map for all objects for which data was available \citep{2019ApJ...887...93G}.

In order to measure the properties of short bursts in CVs we reanalyzed already published \textit{TESS} observations of V1025~Cen \citep{2022ApJ...924L...8L} and TW~Pic \citep{2022NatAs...6...98S}. Moreover, in order to measure individual bursts in MV~Lyr we flux calibrated the $Kepler$ light curve to the $V$ band using calibration of \citet{2017Natur.552..210S}. In the case of TV~Col, EI~UMa and ASASSN-19bh, we corrected the published burst peak luminosities by subtracting the quiescent flux of the systems \citep{2022Natur.604..447S}. The measurements of burst in CP~Pup are directly taken from \citet{Veresvarska}. We supplemented our sample with dwarf novae outbursts reported by \citet{2016MNRAS.460.2526O} that had a duration of seven days or less and were observed by \textit{TESS}.  We note that the constraints on the duration of outbursts exclude most intermediate polars with suspected dwarf nova type outbursts (e.g. GK~Per, V455~And). However, we measured a suspected dwarf nova outbursts in one intermediate polar, FS Aur.

\begin{table*}
\centering
\caption{Properties of short optical bursts observed in CVs together with the reference to the source of their identification. The luminosities and energies are in $TESS$ or $Kepler$ bands.} \label{tab:energies}
\begin{tabular}{ccccccc}
\tablewidth{0pt}
\hline
\hline
		\multirow{ 2}{*}{Object} & \multirow{ 2}{*}{Outburst type} & Peak optical  & Total optical &       Burst      &       Frequency    & \multirow{ 2}{*}{Reference}  \\
		                         &                                 & luminosity [erg/s]         &  energy  [erg]     &     duration [days]    &      [day$^{-1}$]  &   \\
		\hline
		J0333       &  Magnetic gating   & (0.9$\pm$0.5)$\times10^{32}$ & (13.2$\pm$9.8)$\times10^{35}$  &   0.58$\pm$0.18   & 0.13    & 1 \\
		TW Pic      &  Magnetic gating   & (1.5$\pm$0.3)$\times10^{32}$ & (1.9$\pm$1.2)$\times10^{35}$   &   0.05$\pm$0.03   & 16.11   & 2 \\
		MV Lyr      &  Magnetic gating   & (3.0$\pm$0.7)$\times10^{32}$ & (8.7$\pm$4.3)$\times10^{35}$   &   0.12$\pm$0.03   & 6.39    & 3 \\
		V1025 Cen   &  Magnetic gating   & (0.9$\pm$0.1)$\times10^{32}$ & (4.6$\pm$3.5)$\times10^{35}$   &   0.36$\pm$0.06   & 0.35    & 4 \\
		TV Col      &     Micronova      & (0.8$\pm$0.2)$\times10^{34}$ & (1.2$\pm$0.5)$\times10^{38}$   &   0.52$\pm$0.13   & 0.05    & 5\\
		EI UMa      &     Micronova      & (1.9$\pm$0.2)$\times10^{34}$ & (2.6$\pm$0.3)$\times10^{38}$   &   0.36$\pm$0.07   & 0.04    & 5\\
		ASASSN-19bh &     Micronova      & 3.4$\times10^{34}$           & 11.6$\times10^{38}$            &   6.96            & 0.04    & 5\\
		CP Pup      &     Micronova      & (0.32$\pm$0.06)$\times10^{34}$ & 0.6$\times10^{38}$           &   0.8$\pm$0.2     & 0.017   & 6\\
		MQ Dra      &    Donor flare     & 2.6$\times10^{30}$           & 2.2$\times10^{33}$                             &   0.035           & 0.012   & 7\\
	   V1504 Cyg   &    Dwarf nova      & (3.6$\pm$1.3)$\times10^{32}$ & (4.4$\pm$0.3)$\times10^{37}$   &   4.3$\pm$0.2     & 0.09    & 8\\
		IX Dra      &    Dwarf nova      & (2.2$\pm$0.4)$\times10^{32}$ & (2.5$\pm$0.6)$\times10^{37}$   &   3.9$\pm$0.2     & 0.14    & 8\\
		WX Hyi      &    Dwarf nova      & (1.5$\pm$0.2)$\times10^{32}$ & (1.3$\pm$0.3)$\times10^{37}$   &   2.8$\pm$0.4     & 0.18    & 8\\
		SS UMi      &    Dwarf nova      & (8.8$\pm$0.3)$\times10^{31}$ & (8.9$\pm$1.7)$\times10^{36}$   &   3.8$\pm$0.2     & 0.09    & 8\\
		FS Aur      &    Dwarf nova (IP) & (1.3$\pm$0.1)$\times10^{32}$ & (2.0$\pm$0.1)$\times10^{37}$   &   5.7$\pm$0.1     & 0.04    & 8\\
		YZ Cnc      &    Dwarf nova      & (3.7$\pm$0.8)$\times10^{32}$ & (6.8$\pm$1.9)$\times10^{37}$   &   5.8$\pm$1.2     & 0.09    & 8\\
		V485 Cen    &    Dwarf nova      & (4.7$\pm$0.1)$\times10^{31}$ & (4.5$\pm$0.2)$\times10^{36}$   &   3.8$\pm$0.5     & 0.11    & 8\\
	   VW Hyi      &    Dwarf nova      & (2.4$\pm$0.3)$\times10^{32}$ & (3.2$\pm$0.6)$\times10^{37}$   &   6.2$\pm$0.9     & 0.02    & 8\\	   
		X Leo       &    Dwarf nova      & (8.0$\pm$0.1)$\times10^{32}$ & (1.8$\pm$0.1)$\times10^{38}$   &   6.6$\pm$0.3     & 0.09    & 8\\	   	   
		BI Ori      &    Dwarf nova      & (3.7$\pm$0.2)$\times10^{32}$ & (8.5$\pm$0.1)$\times10^{37}$   &   7.0$\pm$0.3     & 0.07    & 8\\	  	   
		AT Cnc      &    Dwarf nova      & (6.9$\pm$0.3)$\times10^{32}$ & (2.1$\pm$0.3)$\times10^{38}$   &   7.7$\pm$0.3     & 0.05    & 8\\	      	   
\hline
\end{tabular}
\textbf{References:} (1) This work; (2) \citet{2022NatAs...6...98S}; (3) \citet{2017Natur.552..210S}; (4) \citet{2022ApJ...924L...8L}; (5) \citet{2022Natur.604..447S}; (6) \citet{Veresvarska}; (7) \citet{2021MNRAS.504.4072R}; (8) \citet{2016MNRAS.460.2526O}
\end{table*}

We compared the bursts in CVs to stellar flares from main sequence stars as they can occur on similar timescales. The only CV with flares originating from the donor observed by $TESS$ is MQ~Dra \citep{2021MNRAS.504.4072R}. MQ~Dra was below the detection limit of ASAS-SN during the $TESS$ observations. Hence, we employed the Zwicky Transient Facility (ZTF; \citealt{2019PASP..131a8003M})  observations in $g$ filter to flux calibrate $TESS$ data. Since main sequence stars can experience superflares that are more energetic than the normal flare in MQ~Dra, we expanded the comparison to superflares in single main sequence stars \citep{2020ApJ...890...46T}. The superflares properties have been transformed to $g$ filter from bolometric values using corrections from \citet{2019A&A...632A.105C}.

In order to estimate the frequency of bursts we follow eq.~9 of \citet{2020ApJ...890...46T}, i.e. the frequency of bursts is equal to the number of bursts divided by the continuous monitoring period. In the case of bursts that were occurring only during a low state of the system, we measured bursts frequency only during the time interval when the bursts were present, rather than the entire monitoring window (i.e. during a low state of MV~Lyr and TW~Pic, see e.g. \citealt{2022NatAs...6...98S}). 

The start of a bursts was assumed to be at a time when the flux of the object visibly rose above the quiescent or noise level. The burst end was assumed to be at a time when the brightness returned to the pre-burst level. We note that in some of the objects the measurements of start and end point of a bursts might have been affected by the changes in brightness due to orbital variability. Therefore, the measured duration of bursts might have a systematic error of up to a few percent.

The collected sample of bursts in CVs is presented in Table~\ref{tab:energies}. The final reported values are the mean of measurements for all of the observed bursts with the reported range corresponding to the largest deviation from the mean of an individual outburst. We note that none of the measured energies or luminosities have bolometric correction. However, they are measured in a consistent fashion, i.e. they are measured in the $TESS$ or $Kepler$ bands and calibrated to either $g$ or $V$ bands.


\section{Results} \label{sec:results}

\begin{figure*}
\begin{flushright}
\includegraphics[width=0.95\hsize]{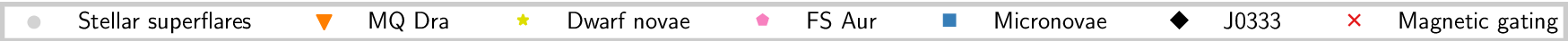}
\end{flushright}
\vspace{-4 mm}
 \centering
\includegraphics[width=0.49\hsize]{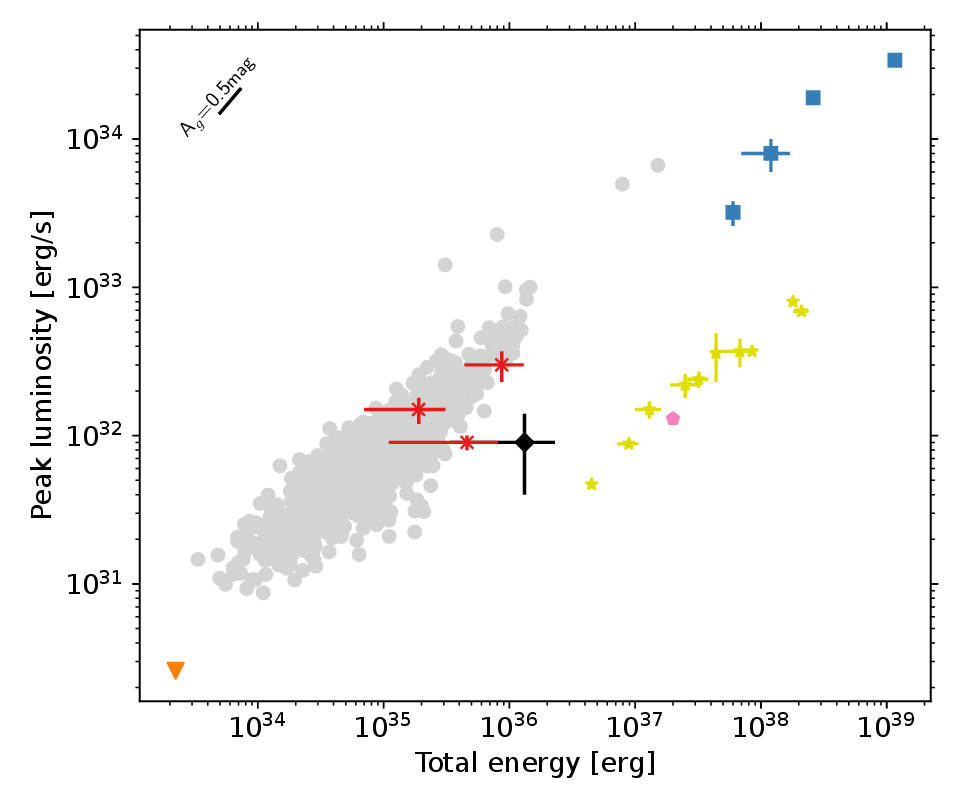}
\includegraphics[width=0.49\hsize]{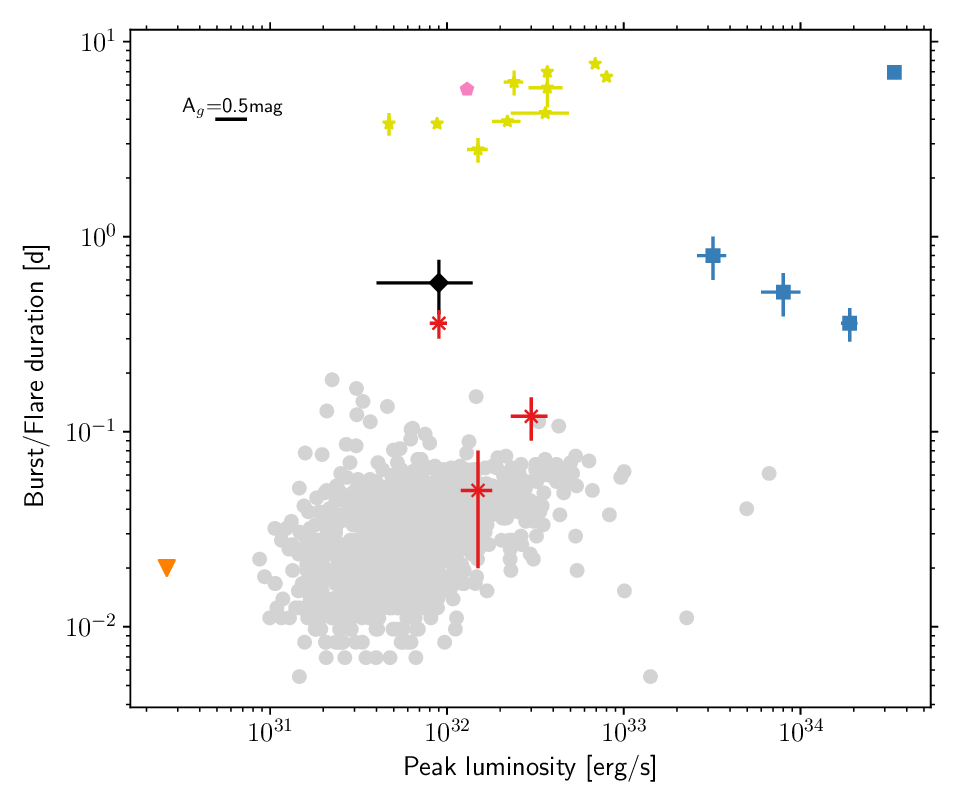}
\includegraphics[width=0.49\hsize]{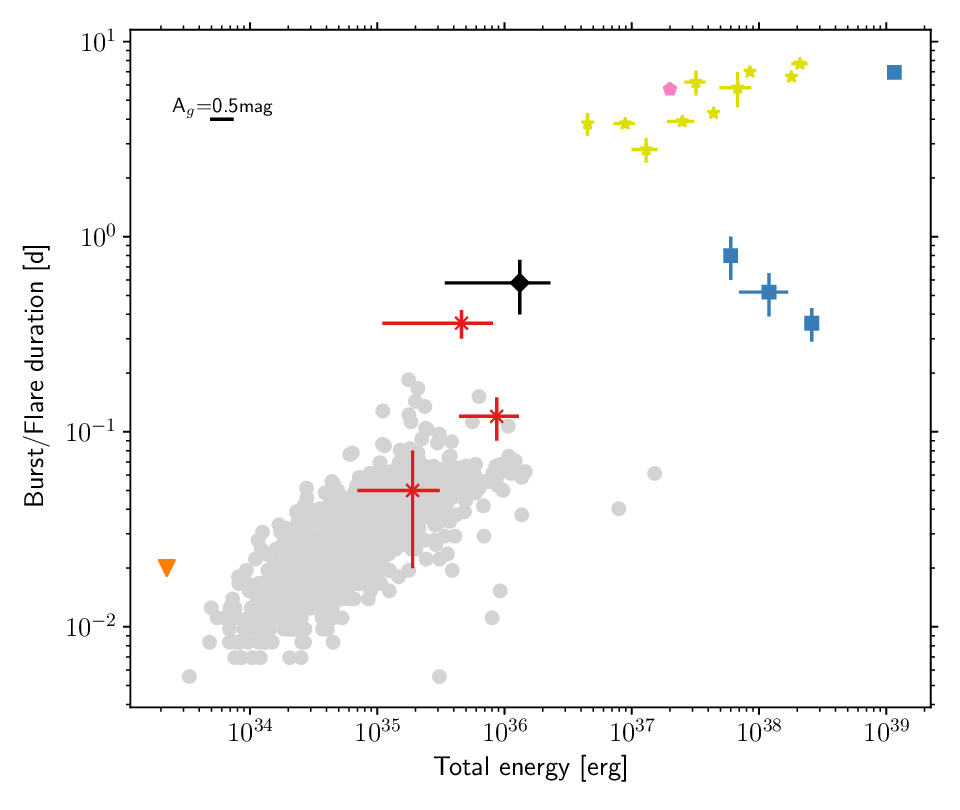}
\includegraphics[width=0.49\hsize]{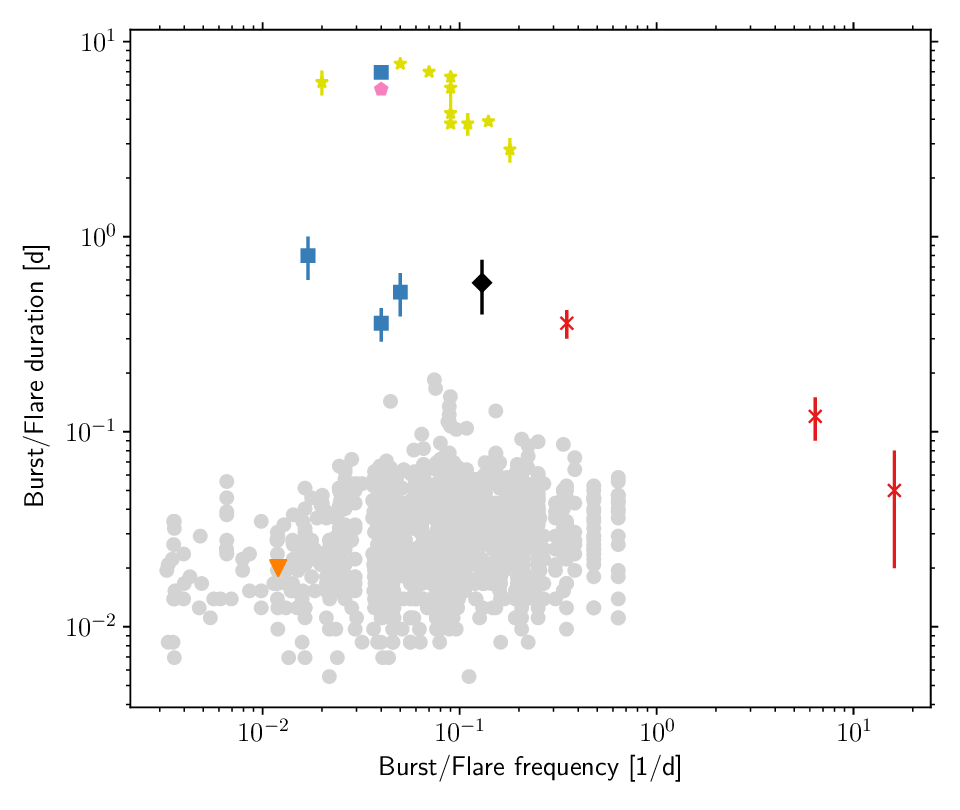}
\caption{Properties of short bursts in CVs with their identification from literature. Superflares from single main sequence stars are plotted for comparison. Effect of extinction of A$_g=0.5$~mag on the measured properties is marked with a black line.}
\label{fig:J0333_diagrams}
\end{figure*}

As a result of the comparison between the bursts we discovered that the bursts properties appear to fall into separate clusters which seem to not be connected. This suggests different physical mechanism behind bursts in each cluster. In fact, these clusters appear to be consistent with the classification suggested in the literature, being  micronovae, magnetic gating bursts, dwarf novae, and donor flares (Fig.~\ref{fig:J0333_diagrams}). We note that \citet{2022A&A...664A...7H} questioned the micronova interpretation and suggested that micronovae can be interpreted as magnetic gating bursts instead. However, it is immediately clear that systems classified as micronovae have energies orders of magnitude higher compared to bursts that were identified as magnetic gating (Table~\ref{tab:energies}). This confirms that magnetic gating and micronovae are two different phenomena \citep{2022Natur.604..447S}. The previous confusion between micronovae and magnetic gating is likely due to the fact that energies of the bursts were not included in the analysis of \citet{2022A&A...664A...7H}.

We propose that the four panels in Fig.~\ref{fig:J0333_diagrams} can be used as diagnostic diagrams that can differentiate between the burst types. A distinction between magnetic gating bursts and stellar flares based solely on energies is not possible. Instead, one has to rely on a relationship between flares/bursts frequency and their average duration (Fig.~\ref{fig:J0333_diagrams}). In particular, there is an apparent anti-correlation between magnetic gating bursts average  duration and their frequency. The anti-correlation between the frequency of magnetic gating bursts and their duration may be expected, as lower burst frequency implies a higher mass that was halted in the accretion disc between the bursts. We note that the micronovae measured frequency is likely significantly overestimated, since they likely remain dormant for significantly longer periods of time compared to their continuous monitoring time. The exception is CP~Pup, where the recurrence time of $\sim$60~days was estimated \citep{Veresvarska}. However, micronovae can be distinguished from other classes of bursts using other properties.

\subsection{Bursts in J0333} \label{sec:bursts}

J0333 was classified as a CV by \citet{2010MNRAS.405..621A}. J0333 showed a blue continuum and emission lines of He~I, He~II, Fe~II, and Balmer series. Based on radial velocity study \citet{2010MNRAS.405..621A} estimated an orbital period of the system to be 0.06~days. However, the orbital period determination was hindered by insufficient sampling of the spectroscopic observations. Moreover, the authors reported strong, short-term variations in the photometric observations of J0333. While \citet{2010MNRAS.405..621A} classified J0333 as a dwarf nova there are no recorded outbursts of the system in the literature.

J0333 experienced six bursts with varying amplitude and duration during the \textit{TESS} monitoring period (Fig.~\ref{fig:J0333_tess}). Majority of bursts appear symmetrical, while some of the larger bursts show a fast rise and a slow decline. Moreover, the first bursts observed displayed a secondary brightening during decline.

\begin{figure}
\includegraphics[width=1.0\hsize]{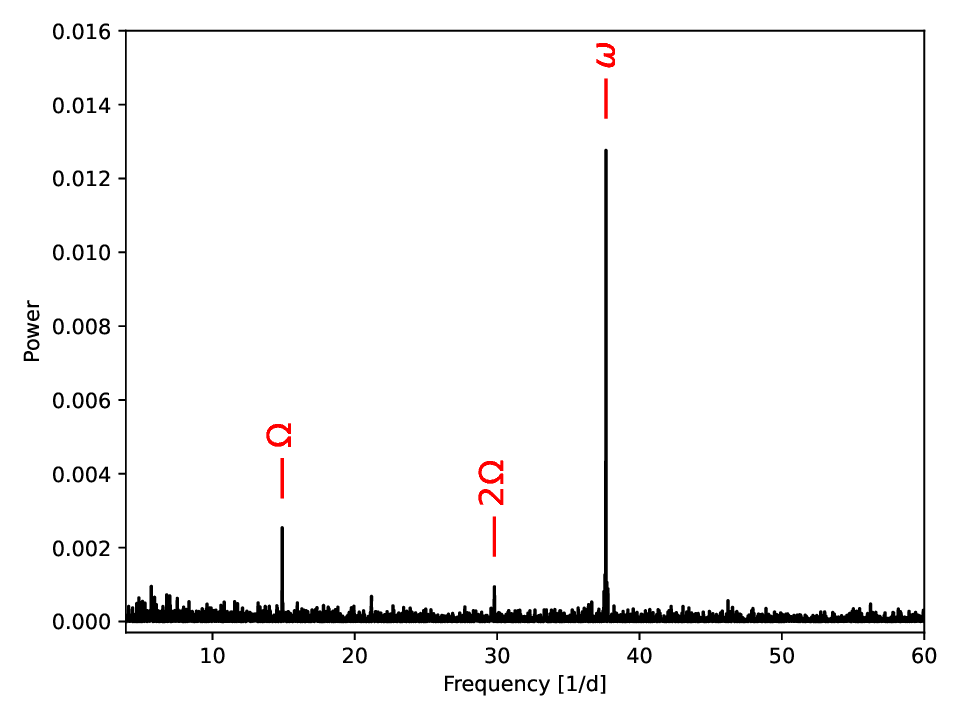}
\caption{Lomb-Scargle periodogram of J0333 $TESS$ observations after the bursts have been masked. Detected periodic variability corresponds to the orbital period ($\Omega$) and the white dwarf spin period ($\omega$). }
\label{fig:J0333_ls}
\end{figure}

After masking the bursts we performed a timing analysis of J0333 $TESS$ data. A Lomb-Scargle periodogram \citep{1976Ap&SS..39..447L,1982ApJ...263..835S} revealed a variability at frequency of 14.89686(6)~d$^{-1}$ (97~min), which we identify as the orbital period (Fig.~\ref{fig:J0333_ls}). This orbital period is consistent with the radial velocity study done by \citet{2010MNRAS.405..621A}. Moreover, we discover variability at a frequency of 37.6422(1)~d$^{-1}$ (38~min). We associate this variability with the white dwarf spin period, suggesting an intermediate polar nature of J0333. The source was detected in X-rays by $Swift-XRT$ and catalogued as 2SXPS J033320.3-445139 \citep{2020ApJS..247...54E}. The $Swift-XRT$ signal-to-noise ratio was insufficient for a detailed timing analysis although hints of a variability at the 38\,min spin period is found. In addition, the X-ray spectrum is rather hard and consistent with an optically thin plasma at a temperature of $\sim$7~keV (Section~\ref{sec:xray_j0333}), further corroborating the intermediate polar identification. However, we note that the J0333 X-ray luminosity of $\sim 10^{32}$~erg/s suggests an unexpectedly high mass transfer rate for a system with a short orbital period.

The properties of J0333 bursts observed by $TESS$ are presented in Table~\ref{tab:energies}. The energies of bursts are similar to both magnetic gating bursts and stellar flares. However, the relatively long duration of  bursts identifies J0333 as a magnetic gating system (Fig.~\ref{fig:J0333_diagrams}).

\section{Discussion and Conclusions } \label{sec:concl}

We showed that short bursts in CVs display distinct observational properties that divide them into separate groups consistent with their literature classification. Based on that we proposed diagnostic diagrams that can distinguish between the short optical bursts types in CVs. The main conclusion from these diagrams regards intermediate polars. Namely, short dwarf nova outbursts in intermediate polars were speculated to be less likely to occur compared to non-magnetic systems \citep{2017A&A...602A.102H}. However, the properties of bursts in an intermediate polar FS~Aur \citep{2013MNRAS.432.2596N} are only consistent with dwarf nova nature of the system (Fig.~\ref{fig:J0333_diagrams}). Moreover, micronovae and magnetic gating intermediate polars are separated on the diagnostic diagrams, contrary to the recent suggestion in the literature \citep{2022A&A...664A...7H}. We note that this does not confirm the physical mechanism suggested for these bursts in the literature, but simply it implies a different physical mechanism are at play in micronovae and magnetic gating systems.

While all micronovae are separated from other classes of short bursts, there is a clear divide between them. Namely, the burst in ASASSN-19bh has a duration and total energy one order of magnitude larger compared to other micronovae. Together with the different shapes of bursts this mimics the two types of Type-I X-ray bursts, as was noted by \citet{2022Natur.604..447S}. However, the comparison to Type-I X-ray bursts is limited due to the fact that the nuclear reactions expected in Type-I X-ray bursts and micronovae differ. Moreover, when a larger sample of systems is discovered, it will be possible to improve the populations in the diagnostic diagrams allowing to confirm or disprove the segregation of different micronovae. Nevertheless, the shared relationship between the total energy released and peak flux during outburst of the currently known sample seem to suggest that they are indeed two classes of the same phenomenon (Fig.~\ref{fig:J0333_diagrams}). If we consider the short duration micronovae alone, it seems that there is an apparent relationship between the burst duration and peak luminosity. While only three such systems are known, a possibility of using short duration micronovae as distance indicators should be investigated when a larger sample of objects will be discovered.

Bursts in intermediate polars have been proposed to be connected to the appearance of superhumps \citep{2023MNRAS.523.3192M}. However, since we have shown that micronovae and magnetic gating bursts seem to be two separate phenomena, it seems that superhumps are only connected to the occurrence of short duration micronovae and do not appear in magnetic gating systems.

We identified J0333 as an intermediate polar with an orbital period of 97~min and a white dwarf spin period of 38~min. Moreover, we derived diagnostic diagrams that identified short bursts discovered in J0333 as a signature of magnetic gating. J0333 has a relatively high spin-to-orbital period ratio of 0.39. The orbital period of J0333 and its spin-to-orbital period ratio is very close to  what is observed in EX~Hya, V598~Peg, V1025~Cen and DW~Cnc \citep[see fig. 5 of ][]{2023ApJ...943L..24L}. Interestingly, EX~Hya, V1025~Cen and DW~Cnc variability has been interpreted as magnetic gating bursts \citep{2007MNRAS.380..353M,2022MNRAS.510.1002D,2022ApJ...924L...8L}. This suggests that magnetic gating in intermediate polars with a high spin-to-orbital period ratio may be common.

\begin{acknowledgments}
This work was supported by Polish National Science Center grant Sonatina 2021/40/C/ST9/00186.  MV acknowledges the support of the Science and Technology Facilities Council (STFC) studentship ST/W507428/1. DdM acknowledges financial support from INAF AstroFund-2022 grant "FANS". AS acknowledges the Warwick Astrophysics PhD prize scholarship made possible thanks to a generous philanthropic donation. This paper includes data collected by the TESS mission. Funding for the TESS mission is provided by the NASA's Science Mission Directorate.
\end{acknowledgments}

\appendix

\section{Long term variability of J03333}\label{sec:long_j0333}

\begin{figure*}[h!]
\includegraphics[width=1.0\hsize]{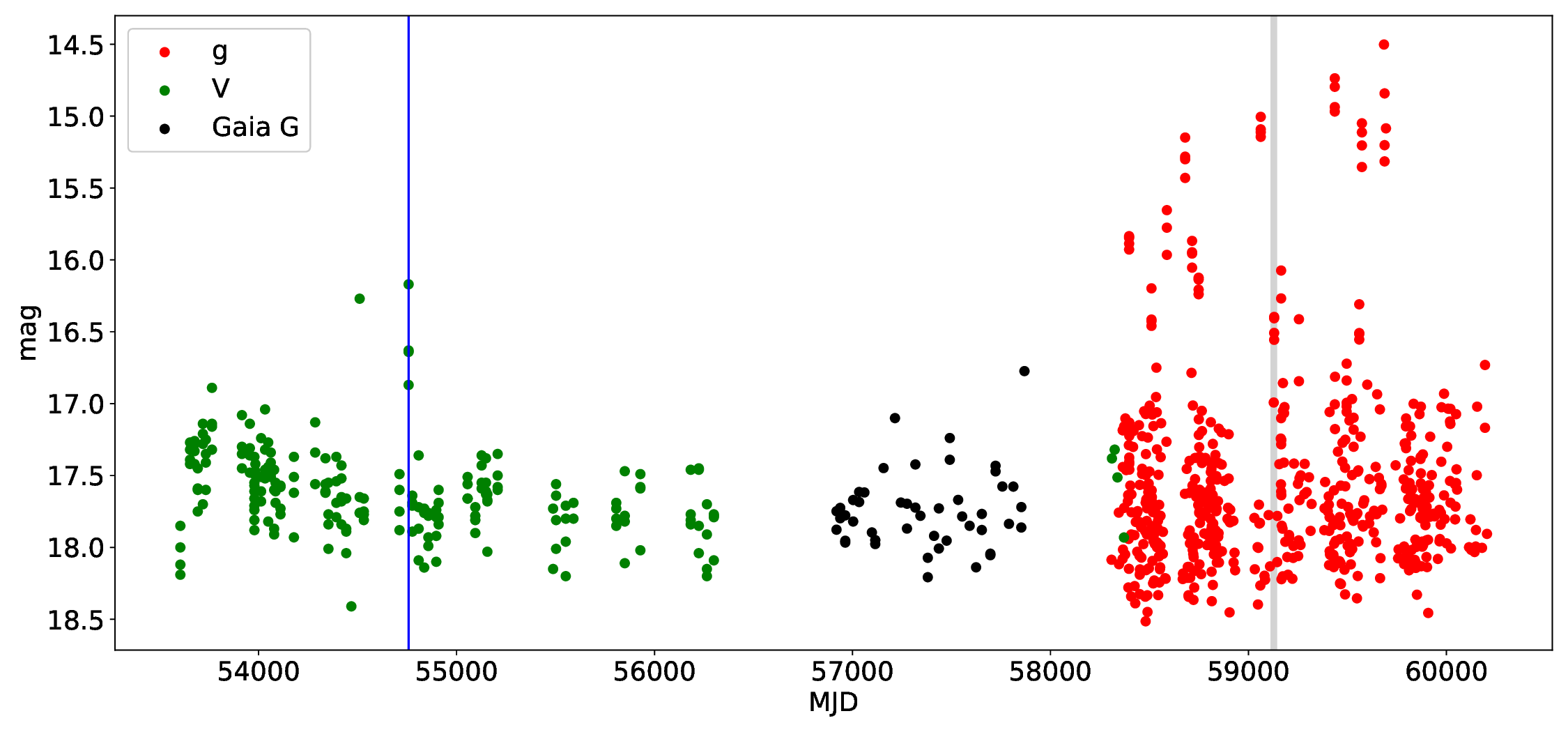}
\caption{A light curve of J0333 in $V$, Gaia $G$ and $g$ filters. The time of \textit{TESS} monitoring is marked with a grey area. Time of the \textit{Swift} pointing is marked with a blue line. The \textit{Swift} pointing  was one day after an apparent brightening in the $V$ band.}
\label{fig:J0333_long_lc}
\end{figure*}

Variability of J0333 on long timescales was analysed using data from the Catalina Real-time Transient Survey \citep[CRTS]{2009ApJ...696..870D}, All-Sky Automated Survey for Supernovae \citep[ASAS-SN]{2014ApJ...788...48S,2017PASP..129j4502K} and Gaia mission \citep{2016A&A...595A...1G,2023A&A...674A..32B}. The long-term variability of J0333 is presented in Figure~\ref{fig:J0333_long_lc}. Similar bursts to what is observed in \textit{TESS} are present during most of the monitoring period in the $g$ band (Fig.~\ref{fig:J0333_long_lc}). Presence of the bursts before that time is inconclusive due to lower cadence of the data, but at least one burst seemed to be observed in the $V$ band one day after the $Swift$ pointing. Both the X-ray flux and UV flux in the $UVM2$ band seemed constant during the $Swift$ pointing, suggesting that the burst was over by the time of $Swift$ observation.

\section{The X-ray emission of J0333}\label{sec:xray_j0333}
J0333 was observed by $Swift-XRT$ on Oct. 19, 2008 for $\sim$21\,h accumulating a total of 10\,ks (ObsID: 00037302001). Is was found at a count rate of 0.05$cts\,s^{-1}$ in the 0.3-10\,keV range. The sparse X-ray coverage due to the spacecraft orbit and the low S/N did not allow a period search but folding the data at the 38\,min period a variability with amplitude of 19$^{+10}_{-9}\%$ was found. The X-ray spectrum averaged over the whole observation is equally fitted (C-statistic) with an absorbed  power law with index 1.84$^{+0.23}_{-0.22}$ ($\chi_{red}^2$=1.14) and an optically thin plasma {\sc apec} with temperature  of 6.6$^{+5.5}_{-1.9}$\,keV adopting solar abundances ($\chi_{red}^2$=1.2) models. The more physical {\sc apec} model fit gives  an hydrogen column density of $\rm N_H=3.8^{+4.4}_{-0.2} \times 10^{20}\,cm^{-2}$ consistent with the small distance of J0333 and the total column density towards the source \citep{2016A&A...594A.116H}, while the power law fit gives a much higher value $\rm N_H=1.2 \times 10^{21}\,cm^{-2}$. The unabsorbed X-ray flux for the {\sc apec} fit was $\rm F_{0.3-10}=1.9\times 10^{-12}\,erg\,cm^{-2}\,s^{-1}$ and the bolometric flux over a dummy range of 0.1-100\,keV resulted in $\rm \sim 2.2\times 10^{-12}\,erg\,cm^{-2}\,s^{-1}$. At the Gaia DR3 distance of 578\,pc the X-ray luminosity was found to be $\rm \sim 8.8 \times 10^{31}\,erg\,s^{-1}$. The lack of better quality data and higher energy coverage prevents to infer a possible temperature  gradient and hence a possible estimate of the white dwarf mass.

\bibliography{references}{}

\begin{thebibliography}{}
\expandafter\ifx\csname natexlab\endcsname\relax\def\natexlab#1{#1}\fi
\providecommand{\url}[1]{\href{#1}{#1}}
\providecommand{\dodoi}[1]{doi:~\href{http://doi.org/#1}{\nolinkurl{#1}}}
\providecommand{\doeprint}[1]{\href{http://ascl.net/#1}{\nolinkurl{http://ascl.net/#1}}}
\providecommand{\doarXiv}[1]{\href{https://arxiv.org/abs/#1}{\nolinkurl{https://arxiv.org/abs/#1}}}

\bibitem[{{Augusteijn} {et~al.}(2010){Augusteijn}, {Tappert}, {Dall}, \&
  {Maza}}]{2010MNRAS.405..621A}
{Augusteijn}, T., {Tappert}, C., {Dall}, T., \& {Maza}, J. 2010, \mnras, 405,
  621, \dodoi{10.1111/j.1365-2966.2010.16487.x}

\bibitem[{{Babusiaux} {et~al.}(2023){Babusiaux}, {Fabricius}, {Khanna},
  {Muraveva}, {Reyl{\'e}}, {Spoto}, {Vallenari}, {Luri}, {Arenou},
  {{\'A}lvarez}, {Anders}, {Antoja}, {Balbinot}, {Barache}, {Bauchet},
  {Bossini}, {Busonero}, {Cantat-Gaudin}, {Carrasco}, {Dafonte}, {Diakit{\'e}},
  {Figueras}, {Garcia-Gutierrez}, {Garofalo}, {Helmi}, {Jim{\'e}nez-Arranz},
  {Jordi}, {Kervella}, {Kostrzewa-Rutkowska}, {Leclerc}, {Licata}, {Manteiga},
  {Masip}, {Mongui{\'o}}, {Ramos}, {Robichon}, {Robin}, {Romero-G{\'o}mez},
  {S{\'a}ez}, {Santove{\~n}a}, {Spina}, {Torralba Elipe}, \&
  {Weiler}}]{2023A&A...674A..32B}
{Babusiaux}, C., {Fabricius}, C., {Khanna}, S., {et~al.} 2023, \aap, 674, A32,
  \dodoi{10.1051/0004-6361/202243790}

\bibitem[{{Bailer-Jones} {et~al.}(2021){Bailer-Jones}, {Rybizki}, {Fouesneau},
  {Demleitner}, \& {Andrae}}]{2021AJ....161..147B}
{Bailer-Jones}, C.~A.~L., {Rybizki}, J., {Fouesneau}, M., {Demleitner}, M., \&
  {Andrae}, R. 2021, \aj, 161, 147, \dodoi{10.3847/1538-3881/abd806}

\bibitem[{{Cannizzo} {et~al.}(2012){Cannizzo}, {Smale}, {Wood}, {Still}, \&
  {Howell}}]{2012ApJ...747..117C}
{Cannizzo}, J.~K., {Smale}, A.~P., {Wood}, M.~A., {Still}, M.~D., \& {Howell},
  S.~B. 2012, \apj, 747, 117, \dodoi{10.1088/0004-637X/747/2/117}

\bibitem[{{Chen} {et~al.}(2019){Chen}, {Girardi}, {Fu}, {Bressan}, {Aringer},
  {Dal Tio}, {Pastorelli}, {Marigo}, {Costa}, \& {Zhang}}]{2019A&A...632A.105C}
{Chen}, Y., {Girardi}, L., {Fu}, X., {et~al.} 2019, \aap, 632, A105,
  \dodoi{10.1051/0004-6361/201936612}

\bibitem[{{D'Angelo} \& {Spruit}(2010)}]{2010MNRAS.406.1208D}
{D'Angelo}, C.~R., \& {Spruit}, H.~C. 2010, \mnras, 406, 1208,
  \dodoi{10.1111/j.1365-2966.2010.16749.x}

\bibitem[{{D'Angelo} \& {Spruit}(2012)}]{2012MNRAS.420..416D}
---. 2012, \mnras, 420, 416, \dodoi{10.1111/j.1365-2966.2011.20046.x}

\bibitem[{{Drake} {et~al.}(2009){Drake}, {Djorgovski}, {Mahabal}, {Beshore},
  {Larson}, {Graham}, {Williams}, {Christensen}, {Catelan}, {Boattini},
  {Gibbs}, {Hill}, \& {Kowalski}}]{2009ApJ...696..870D}
{Drake}, A.~J., {Djorgovski}, S.~G., {Mahabal}, A., {et~al.} 2009, \apj, 696,
  870, \dodoi{10.1088/0004-637X/696/1/870}

\bibitem[{{Duffy} {et~al.}(2022){Duffy}, {Ramsay}, {Steeghs}, {Kennedy},
  {West}, {Wheatley}, {Dhillon}, {Ackley}, {Dyer}, {Galloway}, {Gill}, {Acton},
  {Burleigh}, {Casewell}, {Goad}, {Henderson}, {Tilbrook}, {Str{\o}m}, \&
  {Anderson}}]{2022MNRAS.510.1002D}
{Duffy}, C., {Ramsay}, G., {Steeghs}, D., {et~al.} 2022, \mnras, 510, 1002,
  \dodoi{10.1093/mnras/stab3402}

\bibitem[{{Evans} {et~al.}(2020){Evans}, {Page}, {Osborne}, {Beardmore},
  {Willingale}, {Burrows}, {Kennea}, {Perri}, {Capalbi}, {Tagliaferri}, \&
  {Cenko}}]{2020ApJS..247...54E}
{Evans}, P.~A., {Page}, K.~L., {Osborne}, J.~P., {et~al.} 2020, \apjs, 247, 54,
  \dodoi{10.3847/1538-4365/ab7db9}

\bibitem[{{Gaia Collaboration} {et~al.}(2016){Gaia Collaboration}, {Prusti},
  {de Bruijne}, {Brown}, {Vallenari}, {Babusiaux}, {Bailer-Jones}, {Bastian},
  {Biermann}, {Evans}, {Eyer}, {Jansen}, {Jordi}, {Klioner}, {Lammers},
  {Lindegren}, {Luri}, {Mignard}, {Milligan}, {Panem}, {Poinsignon},
  {Pourbaix}, {Randich}, {Sarri}, {Sartoretti}, {Siddiqui}, {Soubiran},
  {Valette}, {van Leeuwen}, {Walton}, {Aerts}, {Arenou}, {Cropper}, {Drimmel},
  {H{\o}g}, {Katz}, {Lattanzi}, {O'Mullane}, {Grebel}, {Holland}, {Huc},
  {Passot}, {Bramante}, {Cacciari}, {Casta{\~n}eda}, {Chaoul}, {Cheek}, {De
  Angeli}, {Fabricius}, {Guerra}, {Hern{\'a}ndez}, {Jean-Antoine-Piccolo},
  {Masana}, {Messineo}, {Mowlavi}, {Nienartowicz}, {Ord{\'o}{\~n}ez-Blanco},
  {Panuzzo}, {Portell}, {Richards}, {Riello}, {Seabroke}, {Tanga},
  {Th{\'e}venin}, {Torra}, {Els}, {Gracia-Abril}, {Comoretto},
  {Garcia-Reinaldos}, {Lock}, {Mercier}, {Altmann}, {Andrae}, {Astraatmadja},
  {Bellas-Velidis}, {Benson}, {Berthier}, {Blomme}, {Busso}, {Carry},
  {Cellino}, {Clementini}, {Cowell}, {Creevey}, {Cuypers}, {Davidson}, {De
  Ridder}, {de Torres}, {Delchambre}, {Dell'Oro}, {Ducourant}, {Fr{\'e}mat},
  {Garc{\'\i}a-Torres}, {Gosset}, {Halbwachs}, {Hambly}, {Harrison}, {Hauser},
  {Hestroffer}, {Hodgkin}, {Huckle}, {Hutton}, {Jasniewicz}, {Jordan},
  {Kontizas}, {Korn}, {Lanzafame}, {Manteiga}, {Moitinho}, {Muinonen},
  {Osinde}, {Pancino}, {Pauwels}, {Petit}, {Recio-Blanco}, {Robin}, {Sarro},
  {Siopis}, {Smith}, {Smith}, {Sozzetti}, {Thuillot}, {van Reeven}, {Viala},
  {Abbas}, {Abreu Aramburu}, {Accart}, {Aguado}, {Allan}, {Allasia},
  {Altavilla}, {{\'A}lvarez}, {Alves}, {Anderson}, {Andrei}, {Anglada Varela},
  {Antiche}, {Antoja}, {Ant{\'o}n}, {Arcay}, {Atzei}, {Ayache}, {Bach},
  {Baker}, {Balaguer-N{\'u}{\~n}ez}, {Barache}, {Barata}, {Barbier}, {Barblan},
  {Baroni}, {Barrado y Navascu{\'e}s}, {Barros}, {Barstow}, {Becciani},
  {Bellazzini}, {Bellei}, {Bello Garc{\'\i}a}, {Belokurov}, {Bendjoya},
  {Berihuete}, {Bianchi}, {Bienaym{\'e}}, {Billebaud}, {Blagorodnova},
  {Blanco-Cuaresma}, {Boch}, {Bombrun}, {Borrachero}, {Bouquillon}, {Bourda},
  {Bouy}, {Bragaglia}, {Breddels}, {Brouillet}, {Br{\"u}semeister},
  {Bucciarelli}, {Budnik}, {Burgess}, {Burgon}, {Burlacu}, {Busonero}, {Buzzi},
  {Caffau}, {Cambras}, {Campbell}, {Cancelliere}, {Cantat-Gaudin}, {Carlucci},
  {Carrasco}, {Castellani}, {Charlot}, {Charnas}, {Charvet}, {Chassat},
  {Chiavassa}, {Clotet}, {Cocozza}, {Collins}, {Collins}, {Costigan}, {Crifo},
  {Cross}, {Crosta}, {Crowley}, {Dafonte}, {Damerdji}, {Dapergolas}, {David},
  {David}, {De Cat}, {de Felice}, {de Laverny}, {De Luise}, {De March}, {de
  Martino}, {de Souza}, {Debosscher}, {del Pozo}, {Delbo}, {Delgado},
  {Delgado}, {di Marco}, {Di Matteo}, {Diakite}, {Distefano}, {Dolding}, {Dos
  Anjos}, {Drazinos}, {Dur{\'a}n}, {Dzigan}, {Ecale}, {Edvardsson}, {Enke},
  {Erdmann}, {Escolar}, {Espina}, {Evans}, {Eynard Bontemps}, {Fabre},
  {Fabrizio}, {Faigler}, {Falc{\~a}o}, {Farr{\`a}s Casas}, {Faye}, {Federici},
  {Fedorets}, {Fern{\'a}ndez-Hern{\'a}ndez}, {Fernique}, {Fienga}, {Figueras},
  {Filippi}, {Findeisen}, {Fonti}, {Fouesneau}, {Fraile}, {Fraser}, {Fuchs},
  {Furnell}, {Gai}, {Galleti}, {Galluccio}, {Garabato}, {Garc{\'\i}a-Sedano},
  {Gar{\'e}}, {Garofalo}, {Garralda}, {Gavras}, {Gerssen}, {Geyer}, {Gilmore},
  {Girona}, {Giuffrida}, {Gomes}, {Gonz{\'a}lez-Marcos},
  {Gonz{\'a}lez-N{\'u}{\~n}ez}, {Gonz{\'a}lez-Vidal}, {Granvik}, {Guerrier},
  {Guillout}, {Guiraud}, {G{\'u}rpide}, {Guti{\'e}rrez-S{\'a}nchez}, {Guy},
  {Haigron}, {Hatzidimitriou}, {Haywood}, {Heiter}, {Helmi}, {Hobbs},
  {Hofmann}, {Holl}, {Holland}, {Hunt}, {Hypki}, {Icardi}, {Irwin}, {Jevardat
  de Fombelle}, {Jofr{\'e}}, {Jonker}, {Jorissen}, {Julbe}, {Karampelas},
  {Kochoska}, {Kohley}, {Kolenberg}, {Kontizas}, {Koposov}, {Kordopatis},
  {Koubsky}, {Kowalczyk}, {Krone-Martins}, {Kudryashova}, {Kull}, {Bachchan},
  {Lacoste-Seris}, {Lanza}, {Lavigne}, {Le Poncin-Lafitte}, {Lebreton},
  {Lebzelter}, {Leccia}, {Leclerc}, {Lecoeur-Taibi}, {Lemaitre}, {Lenhardt},
  {Leroux}, {Liao}, {Licata}, {Lindstr{\o}m}, {Lister}, {Livanou}, {Lobel},
  {L{\"o}ffler}, {L{\'o}pez}, {Lopez-Lozano}, {Lorenz}, {Loureiro},
  {MacDonald}, {Magalh{\~a}es Fernandes}, {Managau}, {Mann}, {Mantelet},
  {Marchal}, {Marchant}, {Marconi}, {Marie}, {Marinoni}, {Marrese},
  {Marschalk{\'o}}, {Marshall}, {Mart{\'\i}n-Fleitas}, {Martino}, {Mary},
  {Matijevi{\v{c}}}, {Mazeh}, {McMillan}, {Messina}, {Mestre}, {Michalik},
  {Millar}, {Miranda}, {Molina}, {Molinaro}, {Molinaro}, {Moln{\'a}r},
  {Moniez}, {Montegriffo}, {Monteiro}, {Mor}, {Mora}, {Morbidelli}, {Morel},
  {Morgenthaler}, {Morley}, {Morris}, {Mulone}, {Muraveva}, {Musella},
  {Narbonne}, {Nelemans}, {Nicastro}, {Noval}, {Ord{\'e}novic},
  {Ordieres-Mer{\'e}}, {Osborne}, {Pagani}, {Pagano}, {Pailler}, {Palacin},
  {Palaversa}, {Parsons}, {Paulsen}, {Pecoraro}, {Pedrosa}, {Pentik{\"a}inen},
  {Pereira}, {Pichon}, {Piersimoni}, {Pineau}, {Plachy}, {Plum}, {Poujoulet},
  {Pr{\v{s}}a}, {Pulone}, {Ragaini}, {Rago}, {Rambaux}, {Ramos-Lerate},
  {Ranalli}, {Rauw}, {Read}, {Regibo}, {Renk}, {Reyl{\'e}}, {Ribeiro},
  {Rimoldini}, {Ripepi}, {Riva}, {Rixon}, {Roelens}, {Romero-G{\'o}mez},
  {Rowell}, {Royer}, {Rudolph}, {Ruiz-Dern}, {Sadowski}, {Sagrist{\`a}
  Sell{\'e}s}, {Sahlmann}, {Salgado}, {Salguero}, {Sarasso}, {Savietto},
  {Schnorhk}, {Schultheis}, {Sciacca}, {Segol}, {Segovia}, {Segransan},
  {Serpell}, {Shih}, {Smareglia}, {Smart}, {Smith}, {Solano}, {Solitro},
  {Sordo}, {Soria Nieto}, {Souchay}, {Spagna}, {Spoto}, {Stampa}, {Steele},
  {Steidelm{\"u}ller}, {Stephenson}, {Stoev}, {Suess}, {S{\"u}veges}, {Surdej},
  {Szabados}, {Szegedi-Elek}, {Tapiador}, {Taris}, {Tauran}, {Taylor},
  {Teixeira}, {Terrett}, {Tingley}, {Trager}, {Turon}, {Ulla}, {Utrilla},
  {Valentini}, {van Elteren}, {Van Hemelryck}, {van Leeuwen}, {Varadi},
  {Vecchiato}, {Veljanoski}, {Via}, {Vicente}, {Vogt}, {Voss}, {Votruba},
  {Voutsinas}, {Walmsley}, {Weiler}, {Weingrill}, {Werner}, {Wevers},
  {Whitehead}, {Wyrzykowski}, {Yoldas}, {{\v{Z}}erjal}, {Zucker}, {Zurbach},
  {Zwitter}, {Alecu}, {Allen}, {Allende Prieto}, {Amorim},
  {Anglada-Escud{\'e}}, {Arsenijevic}, {Azaz}, {Balm}, {Beck}, {Bernstein},
  {Bigot}, {Bijaoui}, {Blasco}, {Bonfigli}, {Bono}, {Boudreault}, {Bressan},
  {Brown}, {Brunet}, {Bunclark}, {Buonanno}, {Butkevich}, {Carret}, {Carrion},
  {Chemin}, {Ch{\'e}reau}, {Corcione}, {Darmigny}, {de Boer}, {de Teodoro}, {de
  Zeeuw}, {Delle Luche}, {Domingues}, {Dubath}, {Fodor}, {Fr{\'e}zouls},
  {Fries}, {Fustes}, {Fyfe}, {Gallardo}, {Gallegos}, {Gardiol}, {Gebran},
  {Gomboc}, {G{\'o}mez}, {Grux}, {Gueguen}, {Heyrovsky}, {Hoar}, {Iannicola},
  {Isasi Parache}, {Janotto}, {Joliet}, {Jonckheere}, {Keil}, {Kim},
  {Klagyivik}, {Klar}, {Knude}, {Kochukhov}, {Kolka}, {Kos}, {Kutka}, {Lainey},
  {LeBouquin}, {Liu}, {Loreggia}, {Makarov}, {Marseille}, {Martayan},
  {Martinez-Rubi}, {Massart}, {Meynadier}, {Mignot}, {Munari}, {Nguyen},
  {Nordlander}, {Ocvirk}, {O'Flaherty}, {Olias Sanz}, {Ortiz}, {Osorio},
  {Oszkiewicz}, {Ouzounis}, {Palmer}, {Park}, {Pasquato}, {Peltzer}, {Peralta},
  {P{\'e}turaud}, {Pieniluoma}, {Pigozzi}, {Poels}, {Prat}, {Prod'homme},
  {Raison}, {Rebordao}, {Risquez}, {Rocca-Volmerange}, {Rosen}, {Ruiz-Fuertes},
  {Russo}, {Sembay}, {Serraller Vizcaino}, {Short}, {Siebert}, {Silva},
  {Sinachopoulos}, {Slezak}, {Soffel}, {Sosnowska}, {Strai{\v{z}}ys}, {ter
  Linden}, {Terrell}, {Theil}, {Tiede}, {Troisi}, {Tsalmantza}, {Tur},
  {Vaccari}, {Vachier}, {Valles}, {Van Hamme}, {Veltz}, {Virtanen}, {Wallut},
  {Wichmann}, {Wilkinson}, {Ziaeepour}, \& {Zschocke}}]{2016A&A...595A...1G}
{Gaia Collaboration}, {Prusti}, T., {de Bruijne}, J.~H.~J., {et~al.} 2016,
  \aap, 595, A1, \dodoi{10.1051/0004-6361/201629272}

\bibitem[{{Green} {et~al.}(2019){Green}, {Schlafly}, {Zucker}, {Speagle}, \&
  {Finkbeiner}}]{2019ApJ...887...93G}
{Green}, G.~M., {Schlafly}, E., {Zucker}, C., {Speagle}, J.~S., \&
  {Finkbeiner}, D. 2019, \apj, 887, 93, \dodoi{10.3847/1538-4357/ab5362}

\bibitem[{{Hameury} \& {Lasota}(2017)}]{2017A&A...602A.102H}
{Hameury}, J.~M., \& {Lasota}, J.~P. 2017, \aap, 602, A102,
  \dodoi{10.1051/0004-6361/201730760}

\bibitem[{{Hameury} {et~al.}(2022){Hameury}, {Lasota}, \&
  {Shaw}}]{2022A&A...664A...7H}
{Hameury}, J.~M., {Lasota}, J.~P., \& {Shaw}, A.~W. 2022, \aap, 664, A7,
  \dodoi{10.1051/0004-6361/202243727}

\bibitem[{{HI4PI Collaboration} {et~al.}(2016){HI4PI Collaboration}, {Ben
  Bekhti}, {Fl{\"o}er}, {Keller}, {Kerp}, {Lenz}, {Winkel}, {Bailin},
  {Calabretta}, {Dedes}, {Ford}, {Gibson}, {Haud}, {Janowiecki}, {Kalberla},
  {Lockman}, {McClure-Griffiths}, {Murphy}, {Nakanishi}, {Pisano}, \&
  {Staveley-Smith}}]{2016A&A...594A.116H}
{HI4PI Collaboration}, {Ben Bekhti}, N., {Fl{\"o}er}, L., {et~al.} 2016, \aap,
  594, A116, \dodoi{10.1051/0004-6361/201629178}

\bibitem[{{I{\l}kiewicz} {et~al.}(2023){I{\l}kiewicz}, {Miko{\l}ajewska}, \&
  {Stoyanov}}]{2023ApJ...953L...7I}
{I{\l}kiewicz}, K., {Miko{\l}ajewska}, J., \& {Stoyanov}, K.~A. 2023, \apjl,
  953, L7, \dodoi{10.3847/2041-8213/ace9dc}

\bibitem[{{Jenkins} {et~al.}(2016){Jenkins}, {Twicken}, {McCauliff},
  {Campbell}, {Sanderfer}, {Lung}, {Mansouri-Samani}, {Girouard}, {Tenenbaum},
  {Klaus}, {Smith}, {Caldwell}, {Chacon}, {Henze}, {Heiges}, {Latham},
  {Morgan}, {Swade}, {Rinehart}, \& {Vanderspek}}]{2016SPIE.9913E..3EJ}
{Jenkins}, J.~M., {Twicken}, J.~D., {McCauliff}, S., {et~al.} 2016, in Society
  of Photo-Optical Instrumentation Engineers (SPIE) Conference Series, Vol.
  9913, Software and Cyberinfrastructure for Astronomy IV, ed. G.~{Chiozzi} \&
  J.~C. {Guzman}, 99133E, \dodoi{10.1117/12.2233418}

\bibitem[{{Kochanek} {et~al.}(2017){Kochanek}, {Shappee}, {Stanek}, {Holoien},
  {Thompson}, {Prieto}, {Dong}, {Shields}, {Will}, {Britt}, {Perzanowski}, \&
  {Pojma{\'n}ski}}]{2017PASP..129j4502K}
{Kochanek}, C.~S., {Shappee}, B.~J., {Stanek}, K.~Z., {et~al.} 2017, \pasp,
  129, 104502, \dodoi{10.1088/1538-3873/aa80d9}

\bibitem[{{Lasota}(2001)}]{2001NewAR..45..449L}
{Lasota}, J.-P. 2001, \nar, 45, 449, \dodoi{10.1016/S1387-6473(01)00112-9}

\bibitem[{{Littlefield} {et~al.}(2022){Littlefield}, {Lasota}, {Hameury},
  {Scaringi}, {Garnavich}, {Szkody}, {Kennedy}, \&
  {Leichty}}]{2022ApJ...924L...8L}
{Littlefield}, C., {Lasota}, J.-P., {Hameury}, J.-M., {et~al.} 2022, \apjl,
  924, L8, \dodoi{10.3847/2041-8213/ac4262}

\bibitem[{{Littlefield} {et~al.}(2023){Littlefield}, {Mason}, {Garnavich},
  {Szkody}, {Thorstensen}, {Scaringi}, {I{\l}kiewicz}, {Kennedy}, \&
  {Wells}}]{2023ApJ...943L..24L}
{Littlefield}, C., {Mason}, P.~A., {Garnavich}, P., {et~al.} 2023, \apjl, 943,
  L24, \dodoi{10.3847/2041-8213/acaf04}

\bibitem[{{Lomb}(1976)}]{1976Ap&SS..39..447L}
{Lomb}, N.~R. 1976, \apss, 39, 447, \dodoi{10.1007/BF00648343}

\bibitem[{{Masci} {et~al.}(2019){Masci}, {Laher}, {Rusholme}, {Shupe}, {Groom},
  {Surace}, {Jackson}, {Monkewitz}, {Beck}, {Flynn}, {Terek}, {Landry},
  {Hacopians}, {Desai}, {Howell}, {Brooke}, {Imel}, {Wachter}, {Ye}, {Lin},
  {Cenko}, {Cunningham}, {Rebbapragada}, {Bue}, {Miller}, {Mahabal}, {Bellm},
  {Patterson}, {Juri{\'c}}, {Golkhou}, {Ofek}, {Walters}, {Graham}, {Kasliwal},
  {Dekany}, {Kupfer}, {Burdge}, {Cannella}, {Barlow}, {Van Sistine}, {Giomi},
  {Fremling}, {Blagorodnova}, {Levitan}, {Riddle}, {Smith}, {Helou}, {Prince},
  \& {Kulkarni}}]{2019PASP..131a8003M}
{Masci}, F.~J., {Laher}, R.~R., {Rusholme}, B., {et~al.} 2019, \pasp, 131,
  018003, \dodoi{10.1088/1538-3873/aae8ac}

\bibitem[{{Mhlahlo} {et~al.}(2007){Mhlahlo}, {Buckley}, {Dhillon}, {Potter},
  {Warner}, \& {Woudt}}]{2007MNRAS.380..353M}
{Mhlahlo}, N., {Buckley}, D.~A.~H., {Dhillon}, V.~S., {et~al.} 2007, \mnras,
  380, 353, \dodoi{10.1111/j.1365-2966.2007.12077.x}

\bibitem[{{Mukai} \& {Pretorius}(2023)}]{2023MNRAS.523.3192M}
{Mukai}, K., \& {Pretorius}, M.~L. 2023, \mnras, 523, 3192,
  \dodoi{10.1093/mnras/stad1603}

\bibitem[{{Neustroev} {et~al.}(2013){Neustroev}, {Tovmassian}, {Zharikov}, \&
  {Sjoberg}}]{2013MNRAS.432.2596N}
{Neustroev}, V.~V., {Tovmassian}, G.~H., {Zharikov}, S.~V., \& {Sjoberg}, G.
  2013, \mnras, 432, 2596, \dodoi{10.1093/mnras/stt622}

\bibitem[{{Otulakowska-Hypka} {et~al.}(2016){Otulakowska-Hypka}, {Olech}, \&
  {Patterson}}]{2016MNRAS.460.2526O}
{Otulakowska-Hypka}, M., {Olech}, A., \& {Patterson}, J. 2016, \mnras, 460,
  2526, \dodoi{10.1093/mnras/stw1120}

\bibitem[{{Ramsay} {et~al.}(2020){Ramsay}, {Doyle}, \&
  {Doyle}}]{2020MNRAS.497.2320R}
{Ramsay}, G., {Doyle}, J.~G., \& {Doyle}, L. 2020, \mnras, 497, 2320,
  \dodoi{10.1093/mnras/staa2021}

\bibitem[{{Ramsay} {et~al.}(2021){Ramsay}, {Hakala}, \&
  {Wood}}]{2021MNRAS.504.4072R}
{Ramsay}, G., {Hakala}, P., \& {Wood}, M.~A. 2021, \mnras, 504, 4072,
  \dodoi{10.1093/mnras/stab1140}

\bibitem[{{Ricker} {et~al.}(2015){Ricker}, {Winn}, {Vanderspek}, {Latham},
  {Bakos}, {Bean}, {Berta-Thompson}, {Brown}, {Buchhave}, {Butler}, {Butler},
  {Chaplin}, {Charbonneau}, {Christensen-Dalsgaard}, {Clampin}, {Deming},
  {Doty}, {De Lee}, {Dressing}, {Dunham}, {Endl}, {Fressin}, {Ge}, {Henning},
  {Holman}, {Howard}, {Ida}, {Jenkins}, {Jernigan}, {Johnson}, {Kaltenegger},
  {Kawai}, {Kjeldsen}, {Laughlin}, {Levine}, {Lin}, {Lissauer}, {MacQueen},
  {Marcy}, {McCullough}, {Morton}, {Narita}, {Paegert}, {Palle}, {Pepe},
  {Pepper}, {Quirrenbach}, {Rinehart}, {Sasselov}, {Sato}, {Seager},
  {Sozzetti}, {Stassun}, {Sullivan}, {Szentgyorgyi}, {Torres}, {Udry}, \&
  {Villasenor}}]{2015JATIS...1a4003R}
{Ricker}, G.~R., {Winn}, J.~N., {Vanderspek}, R., {et~al.} 2015, Journal of
  Astronomical Telescopes, Instruments, and Systems, 1, 014003,
  \dodoi{10.1117/1.JATIS.1.1.014003}

\bibitem[{{Scargle}(1982)}]{1982ApJ...263..835S}
{Scargle}, J.~D. 1982, \apj, 263, 835, \dodoi{10.1086/160554}

\bibitem[{{Scaringi} {et~al.}(2022{\natexlab{a}}){Scaringi}, {Groot}, {Knigge},
  {Lasota}, {de Martino}, {Cavecchi}, {Buckley}, \&
  {Camisassa}}]{2022MNRAS.514L..11S}
{Scaringi}, S., {Groot}, P.~J., {Knigge}, C., {et~al.} 2022{\natexlab{a}},
  \mnras, 514, L11, \dodoi{10.1093/mnrasl/slac042}

\bibitem[{{Scaringi} {et~al.}(2017){Scaringi}, {Maccarone}, {D'Angelo},
  {Knigge}, \& {Groot}}]{2017Natur.552..210S}
{Scaringi}, S., {Maccarone}, T.~J., {D'Angelo}, C., {Knigge}, C., \& {Groot},
  P.~J. 2017, \nat, 552, 210, \dodoi{10.1038/nature24653}

\bibitem[{{Scaringi} {et~al.}(2022{\natexlab{b}}){Scaringi}, {Groot}, {Knigge},
  {Bird}, {Breedt}, {Buckley}, {Cavecchi}, {Degenaar}, {de Martino}, {Done},
  {Fratta}, {I{\l}kiewicz}, {Koerding}, {Lasota}, {Littlefield}, {Manara},
  {O'Brien}, {Szkody}, \& {Timmes}}]{2022Natur.604..447S}
{Scaringi}, S., {Groot}, P.~J., {Knigge}, C., {et~al.} 2022{\natexlab{b}},
  \nat, 604, 447, \dodoi{10.1038/s41586-022-04495-6}

\bibitem[{{Scaringi} {et~al.}(2022{\natexlab{c}}){Scaringi}, {de Martino},
  {Buckley}, {Groot}, {Knigge}, {Fratta}, {I{\l}kiewicz}, {Littlefield}, \&
  {Papitto}}]{2022NatAs...6...98S}
{Scaringi}, S., {de Martino}, D., {Buckley}, D.~A.~H., {et~al.}
  2022{\natexlab{c}}, Nature Astronomy, 6, 98,
  \dodoi{10.1038/s41550-021-01494-x}

\bibitem[{{Schaefer} {et~al.}(2022){Schaefer}, {Pagnotta}, \&
  {Zoppelt}}]{2022MNRAS.512.1924S}
{Schaefer}, B.~E., {Pagnotta}, A., \& {Zoppelt}, S. 2022, \mnras, 512, 1924,
  \dodoi{10.1093/mnras/stac443}

\bibitem[{{Shappee} {et~al.}(2014){Shappee}, {Prieto}, {Grupe}, {Kochanek},
  {Stanek}, {De Rosa}, {Mathur}, {Zu}, {Peterson}, {Pogge}, {Komossa}, {Im},
  {Jencson}, {Holoien}, {Basu}, {Beacom}, {Szczygie{\l}}, {Brimacombe},
  {Adams}, {Campillay}, {Choi}, {Contreras}, {Dietrich}, {Dubberley},
  {Elphick}, {Foale}, {Giustini}, {Gonzalez}, {Hawkins}, {Howell}, {Hsiao},
  {Koss}, {Leighly}, {Morrell}, {Mudd}, {Mullins}, {Nugent}, {Parrent},
  {Phillips}, {Pojmanski}, {Rosing}, {Ross}, {Sand}, {Terndrup}, {Valenti},
  {Walker}, \& {Yoon}}]{2014ApJ...788...48S}
{Shappee}, B.~J., {Prieto}, J.~L., {Grupe}, D., {et~al.} 2014, \apj, 788, 48,
  \dodoi{10.1088/0004-637X/788/1/48}

\bibitem[{{Sokolovsky} {et~al.}(2023){Sokolovsky}, {Johnson}, {Buson}, {Jean},
  {Cheung}, {Mukai}, {Chomiuk}, {Aydi}, {Molina}, {Kawash}, {Linford},
  {Mioduszewski}, {Rupen}, {Sokoloski}, {Williams}, {Steinberg}, {Vurm},
  {Metzger}, {Page}, {Orio}, {Quimby}, {Shafter}, {Corbett}, {Bolzoni},
  {DeYoung}, {Menzies}, {Romanov}, {Richmond}, {Ulowetz}, {Vanmunster},
  {Williamson}, {Lane}, {Bartnik}, {Bellaver}, {Bruinsma}, {Dugan}, {Fedewa},
  {Gerhard}, {Painter}, {Peterson}, {Rodriguez}, {Smith}, {Sullivan}, \&
  {Watson}}]{2023MNRAS.521.5453S}
{Sokolovsky}, K.~V., {Johnson}, T.~J., {Buson}, S., {et~al.} 2023, \mnras, 521,
  5453, \dodoi{10.1093/mnras/stad887}

\bibitem[{{Tu} {et~al.}(2021){Tu}, {Yang}, {Wang}, \&
  {Wang}}]{2021ApJS..253...35T}
{Tu}, Z.-L., {Yang}, M., {Wang}, H.~F., \& {Wang}, F.~Y. 2021, \apjs, 253, 35,
  \dodoi{10.3847/1538-4365/abda3c}

\bibitem[{{Tu} {et~al.}(2020){Tu}, {Yang}, {Zhang}, \&
  {Wang}}]{2020ApJ...890...46T}
{Tu}, Z.-L., {Yang}, M., {Zhang}, Z.~J., \& {Wang}, F.~Y. 2020, \apj, 890, 46,
  \dodoi{10.3847/1538-4357/ab6606}

\bibitem[{{Veresvarska et al.}(2023)}]{Veresvarska}
{Veresvarska et al.} 2023, \mnras, submitted

\bibitem[{{Warner}(2003)}]{2003cvs..book.....W}
{Warner}, B. 2003, {Cataclysmic Variable Stars},
  \dodoi{10.1017/CBO9780511586491}

\end{thebibliography}
\bibliographystyle{aasjournal}

\end{document}